\documentclass[article,twocolumn,showpacs]{revtex4}

\usepackage{amsmath}
\usepackage{amssymb}
\usepackage{color}
\usepackage{verbatim}
\usepackage{array}
\usepackage{graphicx}
\usepackage{epsfig}
\usepackage{bm}

\newcommand{\ds}{\displaystyle}
\newcommand{\ddsum}[1]{{\displaystyle \sum_{ #1 }}}

\def\bra#1{\mathinner{\langle{#1}|}}
\def\ket#1{\mathinner{|{#1}\rangle}}
\newcommand{\braket}[2]{\langle #1|#2\rangle}

\usepackage{ulem} % for deleting
\def\sprm#1#2{  \left\langle #1 \left\vert \right. #2 \right\rangle   }

\def\mem#1#2#3{  \left\langle #1 \left\vert  #2 \right\vert #3 \right\rangle   }

\begin{document}
\preprint{\hfill\parbox[b]{0.3\hsize}{ }}

%=========Title=======================================================
\title{Quantum correlations in the two--photon decay of few--electron ions}

%=========Authors=====================================================
\author{Filippo Fratini$^{1,2}$\footnote{
E-mail address: fratini@physi.uni-heidelberg.de}, Malte C. Tichy$^{3}$,
Thorsten Jahrsetz$^{1,2}$, Andreas Buchleitner$^{3}$, Stephan Fritzsche$^{2,4}$, Andrey Surzhykov$^{1,2}$}
\affiliation
{\it
$^1$
Physikalisches Institut, Ruprecht-Karls-Universit\"at Heidelberg, Philosophenweg 12, D-69120 Heidelberg, Germany \\
$^2$
GSI Helmholtzzentrum f\"ur Schwerionenforschung, D-64291 Darmstadt, Germany \\
$^3$
Physikalisches Institut, Albert-Ludwigs-Universit\"at Freiburg, Hermann-Herder-Strasse 3, D-79104 Freiburg, Germany\\
$^4$
FIAS Frankfurt Institute for Advanced Studies, D-60438 Frankfurt am Main, Germany
}

%=========Abstract====================================================
\begin{abstract}
A theoretical study of the polarization entanglement of two photons emitted 
in the decay of metastable ionic states is performed within the framework of 
density matrix theory and second--order perturbative approach. Particular 
attention is paid to relativistic and non--dipole effects that become important 
for medium-- and high--$Z$ ions. To analyze these effects, the degree of entanglement 
is evaluated both in the dipole approximation and within the rigorous relativistic theory. 
Detailed calculations are performed for the two--photon $2s_{1/2}\to 1s_{1/2}$ transition 
in hydrogen--like, as well as for the $1s_{1/2}\, 2s_{1/2} \; {}^1S_0 \to 1s_{1/2}^2 \; {}^1S_0$, 
$1s_{1/2} \, 2s_{1/2} \; {}^3S_1\to 1s_{1/2}^2 \; {}^1S_0$ and $1s_{1/2} \, 2p_{1/2} \; {}^3P_0\to 1s_{1/2}^2 \; {}^1S_0$ 
transitions in helium--like ions.
% within both approaches, thereby offering a mean to assess the impact of relativistic-- and non --dipole effects on quantum correlations.}
\end{abstract}

%=========Classification==============================================

%=========Text========================================================
\maketitle

\section{Introduction}

Over the decades,  two--photon bound--bound transitions in atoms and ions have provided a unique 
testing ground for advanced atomic theories. Starting from the early work by G\"oppert--Mayer \cite{GoM31} 
and Breit and Teller \cite{BrT40}, a large number of theoretical studies were carried out to estimate the total 
as well as the energy-- and angle--differential (two--photon) decay rates \cite{Kla69,Au76,TuY84,GoD81,ToL90,SaP98,SuK05,JeS08}. 
When compared with  experimental data \cite{MaS72,MoD04,IlU06,KuT09,TrK10}, these studies revealed important information on relativistic, 
quantum electrodynamics (QED) and many--body phenomena in atomic systems. Besides structure--related investigations, more recent interest  focuses  on the \textit{quantum correlations} between the emitted photons, which can be used to probe  fundamental aspects of modern quantum theory. %{physics
In a series of studies, for example, photon--photon \textit{polarization correlations} were employed to test the 
Bell inequality \cite{AsD82,PiK85,KlD97}. In particular, these investigations demonstrated that the polarization correlations cannot 
be explained by any local realistic theory that uses hidden variables. Hence, together with other Bell test experiments,  
two--photon studies  proved that nature indeed does exhibit quantum--mechanical non--locality.
These results contributed to the long-lasting historical debate of Einstein with Bohr and Schr\"odinger \cite{EiP35,Sch49} 
who introduced the notion of entanglement for denoting nonproduct (pure) states.
% This is well in contrast to Einstein's beliefs %\cite{EiP35,Sch49} in his historical debate with Bohr and Schr\"odinger who introduced the notion of entanglement for denoting nonproduct (pure) states.}

\medskip

In the past, both experimental \cite{PiK85} and theoretical \cite{RaS08,RaS08a} studies of $\gamma - \gamma$ polarization correlation were mainly restricted to the $2s \to 1s$ decay of neutral hydrogen (or deuterium). Much less attention was paid to two--photon transitions in other atomic or ionic species. With the recent advances in heavy--ion accelerator and trap facilities as well as in x--ray detection techniques, new possibilities arise to study spin--correlation phenomena in the decay of heavy, few--electron ions. In the medium-- and high--$Z$ domain, however, a proper analysis of polarization quantum correlations requires detailed knowledge of relativistic and of many--electron effects, as well as of those contributions that arise from the higher-order (non--dipole) terms in the expansion of the electron--photon interaction.

\medskip

In this contribution, we investigate  quantum correlations between the polarization states of two photons emitted in the 
decay of few--electron ions. Most naturally,  spin--correlation phenomena are described within the framework of density matrix theory. 
However, before we present details from this theory, we first summarize  the geometry under which the two--photon decay is considered 
in Section \ref{sec_labeling}. In Section \ref{sub_sec_dm}, then, the general expression for the spin--density matrix of the photon pair 
is derived, in terms of the initial populations of the ionic substates, and of the (second--order) transition amplitudes. The evaluation 
of these amplitudes in relativistic, second--order perturbation theory is thereafter discussed for hydrogen-- and helium--like ions. For 
the latter species, we make use of the independent particle model (IPM) which is appropriate for the analysis of bound--state 
transitions in the high--$Z$ domain \cite{SuJ08, SuV10}. Apart from rigorous relativistic results, we also present simplified expressions 
describing the photons' polarization state within the dipole approximation. These intuitive expressions, derived in Section \ref{sub_sec_NRapprox}, will enable 
us to understand the general behaviour of polarization correlations. In order to provide a quantitative description for these correlations, 
we briefly recall in Section \ref{sec_concurrence} the definition of {\it concurrence} as measure of entanglement. Fully relativistic calculations of the concurrence 
are then performed for the $2s_{1/2} \to 1s_{1/2}$ transition in hydrogen--like, as well 
as $1s_{1/2} \, 2s_{1/2} \; {}^1S_0\to 1s_{1/2}^2 \; {}^1S_0$, $1s_{1/2} \, 2s_{1/2} \; {}^3S_1\to 1s_{1/2}^2 \; {}^1S_0$ 
and $1s_{1/2} \, 2p_{1/2} \; {}^3P_0 \to 1s_{1/2}^2 \; {}^1S_0$ transitions in helium--like ions. Results of these calculations 
are displayed in Section \ref{sec_results}, and are compared to the predictions based on the dipole approximation. From this comparison 
we infer the twofold impact of relativity on  polarization entanglement: apart from (i) the loss of purity of  
the photon states, (ii) the relativistic contraction of the wavefunctions and the non--dipole contributions to the electron--photon interaction 
generally lead to the reduction of concurrence; an effect that becomes prominent for heavy ions and high photon energies. 
A brief summary, together with some perspectives, is given in Section \ref{sec_summary}.

\medskip

Atomics units are used throughout the paper, unless differently stated.

\section{Geometry of the setup and the photon labeling problem}
\label{sec_labeling}

\begin{figure}[t]
\center
\includegraphics[width=.45\textwidth]{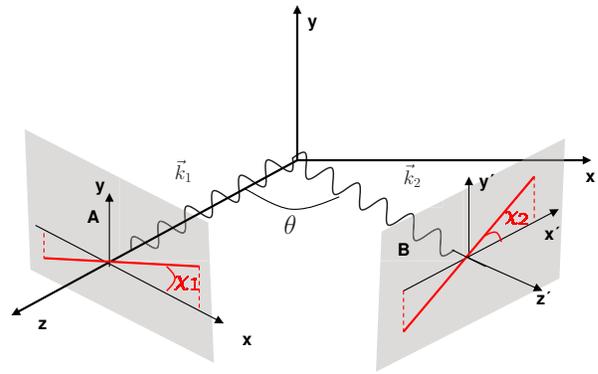}
\caption{(Color online) General decay geometry. The $z$--axis is oriented along the direction of the photon with momentum $\vec k_1$, measured by the detector A. Together with the emission direction of the second photon, it defines  the $x-z$ reaction plane. The opening angle between both photons is denoted by $\theta$, while the (linear) polarization angles of the photons, measured with respect to the $x$--$z$-plane, are denoted by $\chi_1$ and $\chi_2$, respectively.}
\label{geometry}
\end{figure}

In order to analyze  $\gamma-\gamma$ polarization correlations, we  first introduce the geometry of the two--photon emission. Since, for the decay of unpolarized ions, there is no direction \textit{initially preferred} for the overall system, we adopt the momentum of the ``first'' photon to coincide with the $z$ axis which is also taken to be the quantization axis. Together with the direction of the ``second'' photon, this axis defines the reaction plane (x--z plane). A single \textit{opening} angle $\theta$ is required, therefore, to characterize the emission of the photons with respect to each other (cf. Fig.~\ref{geometry}).

\medskip

Since the two photons are in a symmetrized state, it is \textit{a priori} not possible to address them individually. However, we can safely assume that photons observed by the detectors have definite energies and momenta, i.e. they \textit{collapse} onto energy and momentum eigenstates. A clear identity can be given, therefore, to the photons \cite{Tichy}: the {\it first (second)} photon is that one detected by the detector $A(B)$ (marked gray in Fig.~\ref{geometry}) at a certain energy $\omega_{1(2)}$ and with momentum $\vec{k}_{1(2)}$. By distinguishing in such a way the photons, we can use their \textit{polarization states} in order to investigate the associated entanglement properties. Indeed, such an analysis is possible since---in contrast to the energy and momentum spaces---the photon spin state can be directly measured in any basis.

%
%
%
% ------------------------------ Theory -------------------------- %
%
%
\section{Theory}

\subsection{Density matrix approach}
\label{sub_sec_dm}

Having defined the geometry of the two--photon decay, we shall next recall the theoretical background needed to investigate the polarization of the emitted radiation. Most naturally,  polarization--correlation studies can be performed in terms of the system's density matrix. Since this approach was recently applied to describe two--photon transitions in hydrogen--like ions \cite{RaS08,RaS08a}, here we may restrict ourselves to a short compilation of the basic formulas relevant for our further analysis.

\medskip

The initial state of the overall system, in our two--photon decay problem, is given by the photon vacuum $\ket{\rm vac} \equiv \ket{0,0}_{i, \gamma}$, and by the excited ion (or atom) in states $\ket{\alpha_i, J_i,M_i}$ with well--defined total angular momentum $J_i$ and associated projection $M_i$ onto the $z$-axis. Moreover, $\alpha_i$ is a collective label for all additional quantum numbers required for a unique specification of the state. In particular, it characterizes electronic configurations that give rise to the state and, hence, provides its \textit{parity}, $P_i$.

\medskip

The magnetic sublevel population of the ion in initial states is described as a statistical mixture, by the density operator
\begin{equation}
   \hat \rho_{i, {\rm ion}} = \sum_{M_i}C_{M_i}\ket{\alpha_i,J_i,M_i}\bra{\alpha_i,J_i,M_i} \, ,
   \label{initOP}
\end{equation}
where $C_{M_i}$ denotes the population of the magnetic substate $\ket{\alpha_i, J_i, M_i}$. 
Since in most (two--photon) experiments, the initially ``prepared'' excited ionic states are \textit{unpolarized}, we fix the parameters as $C_{M_i} = 1 / (2J_i+1)$. Such a realistic choice for the initial--state population has important consequences for the (spin) entanglement of emitted photon pairs. As we will see later, by introducing the \textit{incoherent} mixture of initial magnetic substates (\ref{initOP}), also the two-photon state's coherences are jeopardized and, hence, a loss of quantum correlations is induced.

\medskip

The density (statistical) operators of the initial and the final states of the overall system are connected by the standard relation \cite{LaK66,CaF67,BaG00}
\begin{equation}
   \hat \rho_f=\hat U \, \hat \rho_{i, {\rm ion}} \otimes \hat{\rho}_{i, \gamma} \, \hat U^{\dag} \, ,
   \label{R}
\end{equation}
where $\hat U$ is the evolution operator which accounts for the interaction of the ion with the radiation field. The final--state operator (\ref{R}) describes both the de--excited ion in some state $\ket{\alpha_f, J_f, M_f}$ and the two emitted photons with momenta $\vec k_{1,2}$ and helicities $\lambda_{1, 2}$. Owing to the transverse character of the electromagnetic radiation, these helicities, or projections of the photon momenta on their own directions of propagation, can take only two values $\lambda_{1, 2} = \pm 1$.

\medskip

Instead of using the final--state density operator $\hat \rho_f$, it is often more convenient to work with its matrix representation, briefly referred to as the final--state \textit{density matrix}. In the representation of the individual angular momenta this matrix reads:
\begin{eqnarray}
   \label{dm_final_general}
   \mem{f; \, \vec k_{1} \lambda_1, \vec k_{2} \lambda_2}{\hat{\rho}_f}{f'; \, \vec k_{1} \lambda'_1, \vec k_{2} \lambda'_2} && \nonumber \\
   && \hspace*{-5.5cm} 
   \equiv \mem{\alpha_f J_f M_f; \, \vec k_{1} \lambda_1, \vec k_{2} \lambda_2}{\hat{\rho}_f}{\alpha_f J_f M'_f; \, \vec k_{1} \lambda'_1, \vec k_{2} \lambda'_2} \nonumber \\
   && \hspace*{-5.5cm} = \frac{1}{2J_i + 1} \sum\limits_{M_i} C_i \mathcal{M}_{fi}^{\vec
   k_1\vec k_2}(\lambda_1,\lambda_2) \mathcal{M}_{fi}^{\vec k_1\vec
   k_2\,*} (\lambda_1',\lambda_2') \, ,
\end{eqnarray}
where we employed Eq.~(\ref{initOP}) in order to evaluate elements of the ionic initial--state matrix $\mem{\alpha_i J_i M_i}{\hat{\rho}_{i, {\rm ion}}}{\alpha_i J_i M'_i}$, and introduced \textit{formal} notation for the transition amplitude:
\begin{eqnarray}
   \label{matrix_element_general}
   \mathcal{M}_{fi}^{\vec k_1\vec k_2}(\lambda_1,\lambda_2) && \nonumber \\
   && \hspace*{-2cm} =
   \mem{\alpha_f J_f M_f; \, \vec k_{1} \lambda_1, \vec k_{2} \lambda_2}{\, \hat{U}}{\alpha_i J_i M_i; 0, 0} \, .
\end{eqnarray}
As seen from this expression, the $\mathcal{M}_{fi}^{\vec k_1\vec k_2}(\lambda_1,\lambda_2)$ describes a transition between two bound ionic states accompanied by the \textit{simultaneous} emission of two photons. 

\medskip

The final--state matrix (\ref{dm_final_general}) still contains the \textit{complete} information about the system and can be used to derive the properties of the photons or the residual ion. Assuming that the final magnetic sub--state of the ion remains unobserved in an experiment, we can derive the \emph{reduced} density matrix $\hat \rho_\gamma$ which only describes the polarization state of the two photons, measured at a certain opening angle $\theta$, with certain energies $\omega_1$ and $\omega_2$:
\begin{equation}
\begin{array}{l}
   \ds\bra{\vec k_1,\lambda_1,\vec k_2,\lambda_2}\hat \rho_{f, \gamma} \ket{\vec k_1,\lambda_1',\vec k_2,\lambda_2'} \\[0.4cm]
   \quad\equiv\ds \sum_{M_f}\bra{f;\vec k_1,\lambda_1,\vec k_2,\lambda_2}\hat\rho_f
   \ket{f;\vec k_1,\lambda_1',\vec k_2,\lambda_2'}\\
   \quad = \ds \frac{{\mathcal N}}{2J_i + 1} \, \sum_{M_i,M_f}C_{M_i} \mathcal{M}_{fi}^{\vec
   k_1\vec k_2}(\lambda_1,\lambda_2) \mathcal{M}_{fi}^{\vec k_1\vec
   k_2\,*} (\lambda_1',\lambda_2') \, , 
\end{array}
\label{dm}
\end{equation}
where we introduced the factor ${\mathcal N}$ to ensure the proper normalization of the matrix, ${\rm Tr}(\hat \rho_\gamma) = 1$. In what follows we will use this (reduced) matrix in order to analyze the polarization entanglement of the photons' pair. Before starting such analysis, we shall briefly discuss the computation of the second--order amplitude (\ref{matrix_element_general}). Most naturally such amplitude can be evaluated within the framework of the second--order perturbation theory \cite{LaK66,CaF67,GoD81}:
\begin{equation}
\begin{array}{l}
   \ds\mathcal{M}_{fi}^{\vec k_1\vec k_2}(\lambda_1,\lambda_2)= \\
   \quad\ds \ddsum{\nu}\!\!\!\!\!\!\!\!\int
   \left[
   \frac{ \bra{f} \hat{\mathcal{R}}(\vec k_1, \lambda_1)
   \ket{\nu}\bra{\nu} \hat{\mathcal{R}}(\vec k_2, \lambda_2) \ket{i}}{E_{\nu}
   -E_i+\omega_2 }  \right.\\
   \quad \left. \ds +\frac{ \bra{f} \hat{\mathcal{R}}(\vec k_2, \lambda_2)
   \ket{\nu}\bra{\nu} \hat{\mathcal{R}}(\vec k_1, \lambda_1) \ket{i}}{E_{\nu} -E_i+\omega_1 } \right] ~ .
   \end{array}
   \label{Mfi}
\end{equation}
Here, $\ket{i}=\ket{\alpha_i,J_i,M_i}$, $\ket{\nu}=\ket{\alpha_\nu, J_\nu, M_\nu}$ and $\ket{f}=\ket{\alpha_f,J_f,M_f}$ denote the solutions of Dirac's equation for the initial, intermediate and final ionic states, respectively, while $E_i$, $E_\nu$ and $E_f$ are the corresponding energies.
Because of energy conservation,  $E_i$ and $E_f$ are related to the energies $\omega_{1,2}$ of the emitted photons by:
\begin{equation}
   E_i - E_f = \omega_{1} + \omega_{2} \, .
\end{equation}
From this relation, it is convenient to define the energy sharing parameter $\eta = \omega_1/(E_i-E_f)$, i.e. the fraction of the energy which is carried away by the ``first'' photon.

\medskip

In Eq.~(\ref{Mfi}) $\hat{\mathcal{R}}(\vec k, \lambda)$ is the transition operator that describes the relativistic interaction of the electrons with the electromagnetic radiation. In velocity (Coulomb) gauge, this operator can be written as a sum of one--particle operators
\begin{eqnarray}
   \label{R_interaction}
   \hat{\mathcal{R}}^{\dag}(\vec k, \lambda) = \sum\limits_{m} \vec\alpha_{m} \vec{\mathcal{A}}_{\lambda, m} = \sum\limits_{m} \vec\alpha_{m} \vec u_{\lambda} e^{i \vec k \cdot\vec r_m} \, ,
\end{eqnarray}
where $\vec\alpha_{m}$ denotes the vector of the Dirac matrices for the $m$--th particle, $\vec{\mathcal{A}}_{\lambda, m}$ is the vector potential of
the photon field and $\vec u_{\lambda}$ is the unit polarization vector. For practical computations, it is convenient to decompose the vector potential $\vec{\mathcal{A}}_{\lambda, m}$ into spherical tensors (i.e., into its electric and magnetic multipole components). For the
emission of the photon in the direction $\hat k = (\theta, \phi)$ with respect to the quantization $z$ axis, such a decomposition reads:
\begin{eqnarray}
   \label{A_decomposition}
   \vec{\mathcal{A}}_{\lambda,m} %index m added
    &=& \sqrt{2 \pi} \sum\limits_{L_\gamma = 1}^{\infty} \sum\limits_{M_\gamma = -L_\gamma}^{L_\gamma} \sum\limits_{p = 0, 1}
   i^{L_\gamma} \, [L_\gamma]^{1/2} \, (i \lambda)^p \, \nonumber \\
   &\times& \hat{a}^{p}_{L_\gamma M_\gamma}(k) \, D^{L_\gamma}_{M_\gamma \lambda}(\hat k) \, ,
\end{eqnarray}
where $[L_\gamma] = 2L_\gamma + 1$, $k=|\vec k|$, $D^{L_\gamma}_{M_\gamma \lambda}$ is the Wigner rotation matrix of rank $L_\gamma$ and the $\hat{a}^{p=0,1}_{L_\gamma M_\gamma}(k)$ refer to magnetic ($p$ = 0) and electric ($p$ = 1) multipoles, respectively. 

\medskip

The great advantage of the multipole expansion (\ref{A_decomposition}), when comparing to the plane--wave formulation in the right--hand--side of Eq.~(\ref{R_interaction}), is that it provides a radial--angular representation of the photon wavefunction. Together with the similar representations of the atomic wavefunctions it allows for significant simplification of the transition amplitude $\mathcal{M}_{fi}^{\vec k_1\vec k_2}(\lambda_1,\lambda_2)$ (see Ref.~\cite{SuK05} for further details). Moreover, Eq.~(\ref{A_decomposition}) gives a very useful tool for studying the multipole effects in the electron--photon interaction. If, for example, the summation in (\ref{A_decomposition}) 
is restricted to the term with $L_\gamma = 1$, $p = 1$, one obtaines the electric--dipole (E1) contribution, while the $L_\gamma = 1$, $p = 0$ 
component provides the magnetic--dipole (M1) contribution, and so on.

\medskip

As seen from Eq.~(\ref{Mfi}), the evaluation of second--order transition amplitudes requires the summation over the \textit{complete} spectrum of the ion. Within the relativistic framework, such a computation is not a simple task since it includes a summation over the discrete part of the Dirac spectrum as well as an integration over the positive-- and negative--energy continua. A number of methods have been developed over the past decades to compute Eq.~(\ref{Mfi}) consistently. Apart from a direct summation over just few intermediate states which are close in energy to the states involved in the decay, the discrete--basis--set approach is widely employed nowadays in (relativistic as well as non--relativistic) second--order calculations. Within this approach, a finite set of discrete pseudostates is constructed from some basis functions and utilized for
computing the transition amplitude $\ds\mathcal{M}_{fi}$ \cite{SaP98,DrG81}. In the present work we use an alternative, {\it Green's--function approach} that helps to avoid the direct summation over the (virtual) intermediate states $\ket{\nu}=\ket{\alpha_\nu, J_\nu, M_\nu}$. In the framework of this alternative approach, moreover, we employ the Sturmian representation \cite{JPB31} of the radial components of the Green's function that allows the \textit{analytical} evaluation of the transition amplitudes (\ref{Mfi}) and, further down, of the entanglement measures.

\medskip

In contrast to hydrogen--like ions, the relativistic second--order calculations for few--electron systems are more intricate, since one has to take into account electron--electron interaction effects. In the high--$Z$ domain, however, the radiative transitions in few--electron ions can be reasonably well understood within the  IPM. This model, which takes the Pauli principle into account, is especially justified for heavy species, since the inter--electronic effects scale with $1/Z$ and, hence, are much weaker than the electron--nucleus interaction. The great advantage of IPM is that it allows the decomposition of the many--body, second--order transition amplitudes in terms of their one--electron analogs (see Ref.~\cite{SuV10} for further details). In Section \ref{sub_sec_helium}, we will apply this approach for the computation of the spin--entanglement between photons emitted in the decay of heavy helium--like ions.

\subsection{Dipole approximation}
\label{sub_sec_NRapprox}

\subsubsection{$S \to S$ transitions}
\label{sub_sub_sect_StoS}

The reduced density matrix (\ref{dm}) contains complete information about the spin states of the photon pairs emitted in the decay of atoms or ions. Together with the transition amplitude (\ref{Mfi}), it is suitable to explore two--photon transitions also in the high--$Z$ domain, where  relativistic and non--dipole effects are significant. However, before we perform such a fully--relativistic analysis, let us restrict ourselves first to the
\textit{non--relativistic dipole} theory and derive approximate expressions for the description of the two--photon polarization states. As we will see later, this will provide intuitive insight into the entanglement properties of the photon pairs. Moreover, by comparing predictions of such
a simplified dipole approach with the results of the fully--relativistic theory, we will be able to identify the relativistic and multipole effects in the two--photon transitions.

\medskip

By making use of the non--relativistic dipole approximation for the electron--photon interaction and by restricting the intermediate--state summation to states $\ket{\nu} = \ket{\alpha_\nu, J_\nu, M_\nu}$ with  definite momentum $J_\nu$ and parity $P_\nu$, defined by dipole selection rules, it is possible to express the two--photon density matrix (\ref{dm}) in the form \cite{RaS08}:
\begin{equation}
\begin{array}{l}
\ds \bra{\vec k_1, \lambda_1,\vec k_2\lambda_2}\hat{\rho}_{\gamma} \ket{\vec k_1, \lambda_1', \vec k_2, \lambda_2'}\\
\quad\ds \approx C \, \lambda_1\lambda_2\lambda_1'\lambda_2'\sum_{L,\mu_1\mu_2}D_{\mu_1\mu_2}^L(\hat x'\hat y'\hat z'\to\hat x\hat y\hat z)\times\\
\quad\times \braket{1,\lambda_1,1,-\lambda_1'}{L,\mu_1}\braket{1,-\lambda_2,1,\lambda_2'}{L,\mu_2}\times\\
\quad\times
\left(
\left\{
\begin{array}{c c c}
J_{\nu}&J_{f}&1\\
J_{i}&J_{\nu}&1\\
1&1&L
\end{array}
\right\}+
\left\{
\begin{array}{c c c}
1&1&L\\
J_{\nu}&J_{\nu}&J_{i}
\end{array}
\right\}
\left\{
\begin{array}{c c c}
1&1&L\\
J_{\nu}&J_{\nu}&J_{f}
\end{array}
\right\}
\right)
~,
\end{array}
\label{DMGen}
\end{equation}
where we employ the standard notation for the Wigner 6$j$-- and 9$j$--symbols \cite{BaG00}, and $C$ is a normalization constant that absorbs the radial parts of the (dipole) transition amplitudes $\mathcal{M}_{fi}$. The final--state density matrix (\ref{DMGen}) depends, therefore, only on the symmetry of the initial and final ionic states as well as on the photons' helicities.
\begin{center}
\begin{figure*}[t]
\includegraphics[width=.6\textwidth]{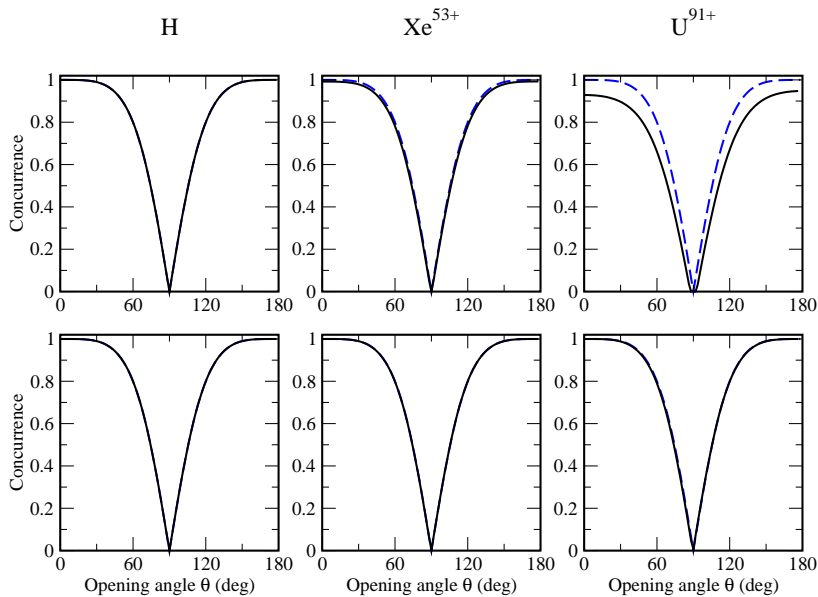}
\caption{(Color online) Concurrence of two photons emitted in the $2s_{1/2} \to 1s_{1/2}$ decay of neutral hydrogen and hydrogen--like xenon and uranium ions. Results of the non--relativistic dipole approximation (dashed line) and the rigorous relativistic theory (solid line) are shown, for two relative photon energies $\eta = 1/16$ (upper panel) and $\eta  = 1/2$ (lower panel).
}
\label{EntSH}
\end{figure*}
\end{center}

\medskip

As mentioned above, Eq.~(\ref{DMGen}) can be applied for the analysis of only those transitions that proceed ---within the non--relativistic picture--- via intermediate (virtual) states $\ket{\alpha_\nu, J_\nu, M_\nu}$ having \textit{one} particular value of total angular momentum $J_\nu$ and parity $P_\nu$.  This is the case of the $1s_{1/2} \, 2s_{1/2} \; {}^1S_0 \to 1s_{1/2}^2 \; {}^1S_0$ transition in helium--like ions, for which the intermediate--state summation in the amplitude (\ref{Mfi}) is restricted to the $n_\nu \, ^1P_1$ levels only, if one treats electron--photon interaction in the dipole approximation. Our approach is also justified for the $2s_{1/2}\to1s_{1/2}$ decay in hydrogen--like ions, since the intermediate states $n_\nu p_{1/2}$ and $n_\nu p_{3/2}$ are degenerate in the non--relativistic limit. In the following analysis, therefore, we shall restrict the discussion of the non--relativistic dipole approximation (\ref{DMGen}) to these two transitions.

\medskip

By inspecting Eq.~(\ref{DMGen}) for the cases of the (non--relativistic) $2s_{1/2} \to 1s_{1/2}$ and $1s_{1/2} \, 2s_{1/2} \; {}^1S_0 \to 1s_{1/2}^2 \; {}^1S_0$ transitions, it can be proven that the density matrix $\hat{\rho}_\gamma$ fulfills the relation ${\rm Tr}(\hat{\rho}^2_\gamma) = ({\rm Tr}(\hat{\rho}_\gamma))^2 = 1$ and, hence, represents a \textit{pure} quantum--mechanical spin state of the emitted photons. This pure state can be described by the state vector:
\begin{equation}
\begin{array}{l}
   \ds \ket{\Psi}= \ds -\frac{1}{2\sqrt{1+\cos^2\theta}}\Big[(\cos\theta-1)(\ket{++} + \ket{--} ) \\
   \quad \ds + (\cos\theta+1)(\ket{+-}+\ket{-+})\Big]  ~ \, ,
\end{array}
\label{six_state}
\end{equation}
as directly derived from Eq.~(\ref{DMGen}). In this expression, the pre--factor arises due to the normalization condition, $\sprm{\Psi}{\Psi} = 1$ and, for the sake of brevity, we use the notation $\ket{\pm} \equiv \ket{\lambda=\pm1}$. Here and henceforth, whenever a state vector describing both photons appears, the first (second) index has to be attributed to the {\it first} ({\it second}) photon; the photons being identified and detected according to Sec.~\ref{sec_labeling}. For the particular case of back--to--back photon emission ($\theta=\pi$), the vector (\ref{six_state}) further simplifies to a Bell state:
\begin{equation}
\begin{array}{l}
   \ds \ket{\Psi}= \frac{1}{\sqrt{2}}(\ket{++}+\ket{--}) \, .
\end{array}
\label{seven_state}
\end{equation}
As seen from this expression, for the opening angle $\theta=\pi$ photons can only be detected having the same helicity. In the past, such a quantum correlation between the photons emitted in the $2s_{1/2} \to 1s_{1/2}$ decay of atomic deuterium has been employed for verifying a violation of Bell's inequality \cite{KlD97}.

\medskip

Instead of the helicity basis $\ket{\lambda=\pm1}$, it is often more convenient to analyze the polarization correlations of two photons in terms of their \textit{linear polarization} unit vectors. These vectors are defined in the plane perpendicular to the photon propagation axis and can be obtained by the standard transformations \cite{Rose}:
\begin{equation}
\begin{array}{l}
   \ket{x}=\frac{1}{\sqrt{2}}\left( \ket{+}+\ket{-} \right) \, ,\\[0.3cm]
   \ket{y}=\frac{i}{\sqrt{2}}\left( -\ket{+}+\ket{-} \right) \, ,
\end{array}
\label{transf}
\end{equation}
and can be used to re--write the state vector (\ref{six_state}) in the $xy$--representation as:
\begin{equation}
\begin{array}{l}
   \ds \ket{\Psi}= -\frac{1}{\sqrt{1+\cos^2\theta}}\Big[\ket{yy}  +
   \cos\theta\ket{xx}\Big] \, .
\end{array}
\label{five_state}
\end{equation}
Since we just change the basis, the opening angle $\theta = \pi$, again corresponds to a Bell, i.e. maximally entangled, state. In contrast, in this representation (\ref{five_state}) we immediately see that perpendicular emission under $\theta = \pi/2$ results in a \textit{product} (non--entangled, or separable) photon spin state.

\medskip

The quantitative analysis of entanglement for the emitted photon pair will be performed in the following Sections, based on our general expression (\ref{DMGen}), as well as on the dipole approximation (\ref{seven_state})--(\ref{five_state}). Before we start with such  analysis, let us still discuss some basic properties of the spin--state (\ref{five_state}). In this way we recall the typical ``experimental scenario'' in which both photons are detected by
polarimeters whose transmission axes are characterized by the angles $\chi_1$ and $\chi_2$ with respect to the reaction (x--z) plane (see Fig.~\ref{geometry}). Eq.~(\ref{five_state}) predicts that after the first photon has been detected by the detector A (the first detector) with some defined (linear) polarization angle $\chi_1$, the second photon \textit{collapses} onto the vector:
\begin{equation}
\begin{array}{l}
   \ket{\Psi}
   \ds \to N\big[ \sin\chi_1\ket{y} + \cos\theta\cos\chi_1\ket{x} \big]
   \end{array} \, ,
\label{eight_state}
\end{equation}
with $N$ some normalization factor. It follows from Eq.~(\ref{eight_state}) that the second photon is then found in a linearly polarized state.
The direction of this linear polarization, characterized by the angle $\tilde{\chi}_2$, depends on the opening angle $\theta$ and on the  polarization angle $\chi_1$ of the first photon:
\begin{equation}
   \tan\tilde{\chi}_2 = \frac{1}{\cos\theta}\tan\chi_1 \, .
   \label{relation}
\end{equation}
As we will show later, such a definite (except for the opening angle $\theta = \pi/2$) correspondence between the linear polarizations does not generally imply maximal entanglement of the photon pairs. To understand this issue and to quantify the degree of entanglement we shall introduce, in Section \ref{sec_concurrence}, the concurrence measure.

\subsubsection{$P \to S$ transitions}
\label{sub_sub_sect_StoP}

In contrast to the $S \to S$ transitions from above, the non--relativistic dipole approximation (\ref{DMGen})--(\ref{five_state}) \textit{cannot} be applied for the analysis of the $1s_{1/2} \, 2p_{1/2} \; {}^3P_0 \to 1s_{1/2}^2 \; {}^1S_0$ decay of helium--like ions. The principal reason for this failure is that the leading (electric--magnetic dipole) E1M1--M1E1 $1s_{1/2} \, 2p_{1/2} \; {}^3P_0 \to 1s_{1/2}^2 \; {}^1S_0$ transition may proceed either via intermediate $1s_{1/2} \, n_\nu s_{1/2} \; {}^3S_1$ or $1s_{1/2} \, n_\nu p \; {}^3P_1$ states, thus giving rise to a ``double-slit'' picture. By taking into account such a Young--type interference and by restricting ourselves to the dipole (E1M1--M1E1) terms in the electron--photon interaction, we again find the photon pair in a pure state:
\begin{equation}
   \begin{array}{l l l}
   \ds \ket{\Psi}&=&\ds C\Big(
   -\Sigma(\eta) \sin^2\frac{\theta}{2}\ket{++}+\Delta(\eta)\cos^2\frac{\theta}{2}\ket{-+}\\
   &&-\Delta(\eta)\cos^2\frac{\theta}{2}\ket{+-}+
   \Sigma(\eta)\sin^2\frac{\theta}{2}\ket{--} \Big) \, .
   \end{array}
   \label{ketP}
\end{equation}
Here, $C$ is the normalization constant, and the energy--dependent functions
$\Sigma(\eta) = S_{E1M1}(\omega_1) + S_{E1M1}(\omega_2) +S_{M1E1}(\omega_1)+S_{M1E1}(\omega_2)$ and $\Delta(\eta) = S_{E1M1}(\omega_1)-S_{E1M1}(\omega_2) -S_{M1E1}(\omega_1) + S_{M1E1}(\omega_2)$ are given in terms of the multipole, second-order reduced transition amplitudes $S_{L_1 p_1, L_2 p_2}(\omega)$ (see Ref.~\cite{SuV10} for further details).

\medskip

As seen from Eq.~(\ref{ketP}), the spin--state of the photons emitted in the $1s_{1/2} \, 2p_{1/2} \; {}^3P_0 \to 1s_{1/2}^2 \; {}^1S_0$ transition depends on the energy sharing $\eta$. No simple analytical expression for this dependence can be derived in general, owing to the complicated structure of the functions $\Delta(\eta)$ and $\Sigma(\eta)$. However, if both photons carry away the same fraction of the energy, $\omega_1 = \omega_2$, the function $\Delta(\eta = 0.5)$ vanishes, and the vector (\ref{ketP}) represents a maximally entangled (Bell) state:
\begin{equation}
   \ds \ket{\Psi}= \frac{1}{\sqrt{2}}\Big(-\ket{++}+\ket{--} \Big)=
    -\frac{i}{\sqrt{2}}\Big(\ket{xy}+\ket{yx} \Big) \, .
\label{kketP}
\end{equation}
By comparing this expression with Eq.~(\ref{five_state}), one can see that polarization properties of $P_0 \to S_0$ and $S_0 \to S_0$ (as well as non--relativistic $s_{1/2} \to s_{1/2}$) transitions  are rather different: while the photons emitted in the $S \to S$ transitions can be detected having {\it parallel} linear polarization vectors, the $P \to S$ decay should result in emission of the photon pair with {\it orthogonal} linear polarizations. Moreover, no angular dependence arises in the state vector (\ref{kketP}) implying maximal entanglement between the photons' spins, irrespective of the particular decay geometry.

\section{Entanglement of the two-photon state}
\label{sec_concurrence}

We are ready now to discuss the concept of entanglement for the emitted photon pairs and to introduce a proper measure for it. In this way, let us first return to the \textit{full} two--photon state which accounts not only for the spin but also for the spatial degrees of freedom.  Within the non--relativistic dipole approximation (\ref{five_state}), such a state \emph{before} its detection reads:
\begin{eqnarray}
   \ket{\Phi}&=&N \int \mbox{d} \omega_1 \mbox{d} \omega_2 \delta(\omega_1+\omega_2-\Delta_\omega) f(\omega_1) \ket{\omega_1\omega_2}
   \nonumber \\ &\times &
   \int \mbox{d} \theta_1 \mbox{d} \theta_2 \ket{\theta_1\theta_2} \Big( \ket{yy} \nonumber \\
   & +&  \cos(\theta_1-\theta_2) \ket{xx} \Big) +(1\leftrightarrow 2) \, ,
\label{full_state}
\end{eqnarray}
where $(1 \leftrightarrow 2)$ denotes the previous terms but with all particles' labels exchanged and the state is normalized by virtue of the constant $N$. Moreover, $f(\omega_{1,2})$ is the energy probability density function of the decay, $\Delta_{\omega} = E_i - E_f$ is the transition energy and $\theta_{1,2}$ are the angles which address the position of the first and second photon in the reaction plane, respectively. Due to the integral over angles and energies, the above state can be written as product state neither in the energies nor in the emission angles or in the polarization. It can hence be seen as highly entangled, in general.

\medskip

In order to rigorously discuss entanglement, due to the identity of particles, a degree of freedom for discrimination is needed. The energy of the photons and their opening angles are an appropriate choice, since one naturally projects onto energy and momentum eigenstates in the experiment. In the coincidence experiment displayed in Fig.~\ref{geometry}, the two--photon state collapses onto a state with definite  momenta, while the polarization can be measured in any basis. If the energies of the photons are equal and the emission directions exactly the same, we are not able to identify two separated particles between which entanglement may be defined. As long as this is not the case, we identify the particle projected on the two energies and angles as the two distinct entities to which we can safely assign an entanglement measure \cite{Tichy}. Hence, even though we start with a rather complex state of identical particles in which no physical subsystem structure is apparent, we can effectively deal with the two--qubit system of polarized photons projected on energy and momentum states.

\medskip

Having clarified the concept of the two--photon entanglement, we shall introduce now its quantitative measure. For a photon pair, that can be seen as a ``two--qubit'' system, it is very convenient to describe the degree of entanglement by means of the Wootter's concurrence $\mathcal{C}$ \cite{PRL80}. For an arbitrary two--qubit system described by the density operator $\hat{\rho}$ the concurrence is defined as
\begin{equation}
   \mathcal{C} = {\rm max} \Big(0,\sqrt{e_1}-\sqrt{e_2}-\sqrt{e_3}-\sqrt{e_4} \Big) \, ,
\label{concDM}
\end{equation}
where $\sqrt{e_i}$ are the square roots of the eigenvalues of the operator $\hat{\rho} (\hat{\sigma}^{(1)}_2 \otimes \hat{\sigma}^{(2)}_2) \hat{\rho}^*(\hat{\sigma_2}^{(1)} \otimes \hat{\sigma}^{(2)}_2)$ in descending order, $\hat{\rho}^*$ is the complex conjugate of $\hat{\rho}$, and $\hat{\sigma}^{(1,2)}_2 = \hat{\sigma}^{(1,2)}_y$ are the Pauli matrices acting on the first and the second qubit, respectively.
Before we discuss further the properties of the concurrence $\mathcal{C}$, let us first recall that it quantifies correlations that can be \textit{fully} attributed to the entanglement. Bi--particle states with vanishing concurrence can still exhibit correlations which are, however, not of quantum nature.

\medskip

Definition (\ref{concDM}) can be simplified further if applied to a pure quantum--mechanical state described by a ket vector
\begin{equation}
\begin{array}{l}
   \ket{\beta}=C_{aa}\ket{aa}+C_{ab}\ket{ab} +\\
   \qquad\qquad C_{ba}\ket{ba} +C_{bb}\ket{bb} ~,
\end{array}
\end{equation}
where $a,b$ are arbitrary two--dimensional basis states 
and $C_{ij}$ are complex numbers. For this state, the concurrence reads
\begin{equation}
   \mathcal{C}=2\Big|C_{aa}C_{bb}-C_{ab}C_{ba}  \Big| \, .
\label{concketgen}
\end{equation}
By using this expression and Eq.~(\ref{five_state}), we immediately obtain the analytical expression
\begin{equation}
\mathcal{C}(\theta)=2\frac{|\cos\theta|}{1+\cos^2\theta} \,
\label{concket}
\end{equation}
for the spin--entanglement of the photons emitted in the $2s_{1/2}\to1s_{1/2}$ and $1s_{1/2} \, 2s_{1/2} \; {}^1S_0\to1s_{1/2}^2 \; {}^1S_0$ transitions. We remind that Eq.~(\ref{concket}) is obtained within the non--relativistic dipole approximation and should be questioned in the high--$Z$ domain, where higher--order and relativistic effects can play a significant role. To explore the influence of these effects on the photon spin--entanglement, we will compare in the next section  the predictions obtained from Eq.~(\ref{concket}) with the rigorous relativistic calculations based on Eqs.~(\ref{dm})--(\ref{Mfi}) and (\ref{concDM}).

\begin{center}
\begin{figure}[t]
\includegraphics[width=.3\textwidth]{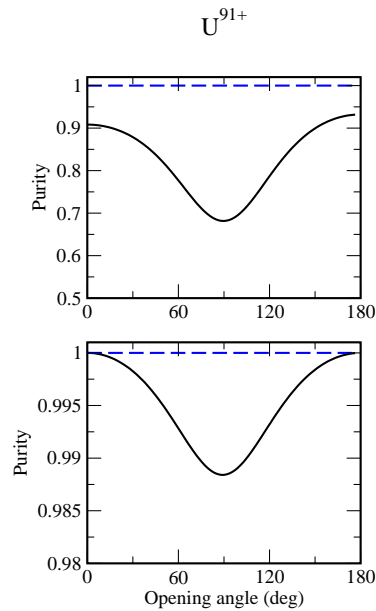}
\caption{(Color online) Purity (\ref{purityDef}) of the two-photon state in the $2s_{1/2} \to 1s_{1/2}$ decay of hydrogen--like uranium. Results of the non--relativistic dipole approximation (dashed line) and a rigorous relativistic treatment (solid line) are shown, for relative photon energies $\eta = 1/16$ (upper panel) and $\eta = 1/2$ (lower panel).
}
\label{purity}
\end{figure}
\end{center}
\begin{center}
\begin{figure}[t]
\includegraphics[width=.5\textwidth]{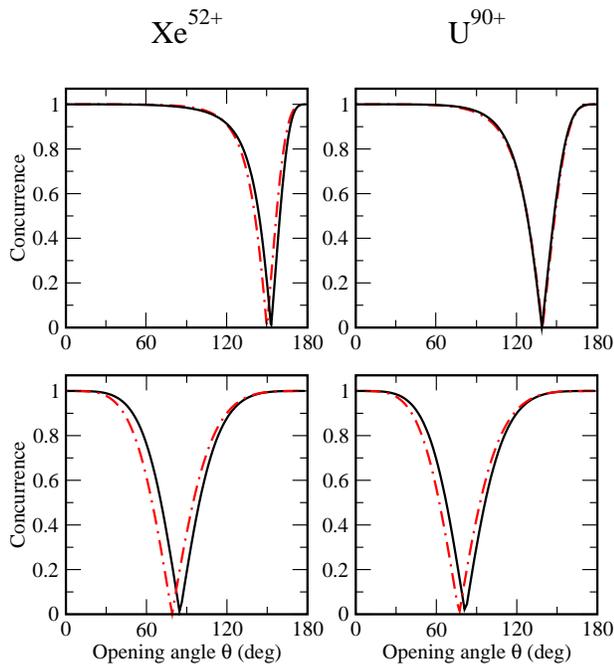}
\caption{(Color online) Concurrence of two photons emitted in the $1s_{1/2} \, 2p_{1/2} \; {}^3P_0 \to 1s_{1/2}^2 \; {}^1S_0$ decay of helium--like xenon and uranium ions. Results of the $E1M1$ dipole approximation (dash-dotted line) and of rigorous relativistic theory (solid line) are shown, for two relative photon energies $\eta = 1/10$ (upper panel) and $\eta = 1/4$ (lower panel).
}
\label{Fig4}
\end{figure}
\end{center}

\section{Results and discussion}
\label{sec_results}
\subsection{Hydrogen--like ions}
\label{sub_sec_hydrogen}

\subsubsection{Polarization entanglement}

After discussing the theoretical background of  two--photon polarization studies, we are prepared now to analyze the influence of the relativistic and higher--multipole effects on  quantum correlations between the emitted particles. We start our analysis with the $2s_{1/2}\to 1s_{1/2}$ decay of hydrogen--like ions, that is well established both in theory and in experiment. As shown above, the polarization properties of this transition can be described---within the non--relativistic dipole approximation---by the state vector (\ref{five_state}) and, hence, by the degree of entanglement (\ref{concket}). The theoretical predictions, obtained within such a non--relativistic approach, are displayed in Fig.~\ref{EntSH} for the decay of neutral hydrogen as well as hydrogen--like xenon Xe$^{53+}$ and uranium U$^{91+}$ ions and are compared with the results of the rigorous relativistic treatment. For the latter, one deals with the relativistic Dirac's wave--functions and includes, in addition, the full interaction between the electron and the radiation field in the amplitude (\ref{Mfi}). Relativistic as well as non--relativistic dipole calculations of the concurrence (\ref{concDM})--(\ref{concketgen}) are performed at two relative photon energies $\eta$ = 1/16 (upper panel) and $\eta$=1/2 (lower panel). As seen from the figure, in case of equal energy sharing ($\eta$ = 1/2), both approaches yield almost identical results along the entire isoelectronic sequence. Our calculations show that, while being maximal for the parallel ($\theta$ = 0) and back--to--back ($\theta = \pi$) photon emission, the concurrence vanishes at the opening angle $\theta = \pi/2$. This behaviour is well understood from Eqs.~(\ref{six_state}) and (\ref{five_state}) as well as from the conservation of the projection $M_{\rm tot}$
of the total angular momentum $J_{\rm tot}$ of the overall system ``ion + two photons''. Namely, if no electron--spin flip were to occur during the $2s_{1/2} \to 1s_{1/2}$ decay and assuming $0$ nuclear spin, the conservation law enforces that the change of the projection of the ion's total angular momentum relative to the quantization axis $z$ (chosen along the momentum of the first photon) would be given by $M_i - M_f = 0 = \lambda_1 + M_{\gamma_2}$. In this expression, $\lambda_1$ is the helicity of the first photon and $M_{\gamma_2}$ is the projection of the angular momentum of the second one. For photons emitted in parallel or back--to--back, this projection is $M_{\gamma_2} = \lambda_2$ and $M_{\gamma_2} = - \lambda_2$, correspondingly, thus leading to the conditions $\lambda_1 = - \lambda_2$ or $\lambda_1 = \lambda_2$. Moreover, owing to the spherical symmetry of $s$--ionic states there is an equal probability of emission of the ``first'' photon with helicity $\lambda_1 = + 1$ or $-1$. This immediately implies maximally entangled Bell states $\ket{\Psi} = (\ket{+-} + \ket{+-})/\sqrt{2}$ for $\theta = 0$, and $\ket{\Psi} = (\ket{++} + \ket{--})/\sqrt{2}$ for $\theta = \pi$, as predicted by Eqs.~(\ref{six_state}) and (\ref{seven_state}).

\medskip

Similar to the cases of parallel and back--to--back photon emission, the conservation condition $\lambda_1 = - M_{\gamma_2}$ with the helicity of the first photon being $\lambda_1 = \pm 1$ may help to understand the behaviour of the entanglement measure $C(\theta)$ at $\theta= \pi/2$. This will require us to return to Eq.~(\ref{A_decomposition}) which simplifies---within the dipole approximation---to:
\begin{equation}
   \vec{\mathcal{A}}_{\lambda} = -\sqrt{6 \pi} \sum\limits_{M_\gamma =-1}^{1}
   \lambda \,  \hat{a}^{p=1}_{1 M_\gamma}(k)  \,  d^{1}_{M_\gamma \lambda}(\theta) \, ,
   \label{A_Pi2}
\end{equation}
where $d^{1}_{M_\gamma \lambda}(\theta)$ is the Wigner's (small) $d$--matrix whose properties are discussed in detail in~\cite{BaG00}. For the  opening angle $\theta = \pi/2$, the elements of this matrix are: $d^{1}_{1 1} = d^{1}_{1 -1} = d^{1}_{-1 1} = d^{1}_{-1 -1} = 1/2$, implying, together with Eq.~(\ref{Mfi}) and with the fact that the ionic states are spherically symmetric, that the probability for the second photon to have projection $M_{\gamma_2} = \mp 1$ on the quantization axis of the overall system is independent of its helicity $\lambda_2$. Thus, no correlations between the polarization (spin) states of the emitted photons appear for the perpendicular emission, leading to the vanishing entanglement $C(\pi/2) = 0$ as displayed in Fig.~\ref{EntSH}.

\medskip

\subsubsection{Purity of the two--photon state and impact on entanglement}

As seen from the top panel of Fig.~\ref{EntSH}, the accuracy of the non--relativistic approximation (\ref{concket}) becomes generally worse if one of the photons has a significantly larger energy than the other one. For the $2s_{1/2} \to 1s_{1/2}$ transition of hydrogen--like uranium, for example, the non--relativistic dipole approximation overestimates the concurrence measure by about $10 \%$ for forward as well as backward opening angles,  for an energy sharing $\eta$ = 1/16. In order to understand better such an energy-dependent behavior, we  study the {\it purity} of the two--photon polarization state, defined as
\begin{equation}
   \mathcal{P}=\frac{4}{3} \, \textrm{Tr}[\hat{\rho}_\gamma^2] - \frac{1}{3} \, ,
   \label{purityDef}
\end{equation}
where $\hat{\rho}_\gamma$ represents the photon density matrix (\ref{DMGen}). The purity varies from 0 (completely mixed state) to 1 (pure state). In Fig.~\ref{purity} we display the purity $\mathcal{P}$ for the $2s_{1/2}\to 1s_{1/2}$ decay of U$^{91+}$ for two relative photon energies: $\eta$ = 1/16 (upper panel) and 1/2 (lower panel). As seen from the figure, the purity strongly depends on the energy sharing parameter: 
while the purity of the two-photon state is always $>$ 0.987 for an equal energy sharing $\eta$ = 0.5, 
it is significantly reduced for $\eta$ = 1/16. The loss of purity can be attributed to the spin--orbit coupling in hydrogen--like ions as well as to the magnetic terms in electron--photon interaction. Both these relativistic effects increase  with the nuclear charge $Z$ and with the photon energy $\omega$. They lead to the 
fact that 
%the $2s_{1/2}\to 1s_{1/2}$ transition \textit{cannot} be viewed anymore as a transition between two pure $2s$ and $1s$ states. Consequently, 
the decay of the unpolarized and, hence, mixed $2s_{1/2}$ level results in the emission of photons characterized by a partially mixed state. Due to complementarity of entanglement and mixedness/impurity \cite{Bergou}, such a loss of purity causes the \textit{reduction} of the concurrence measure that can be observed in the top panel of Fig.~\ref{EntSH}. Despite such a reduction, there are still quantum correlations between the polarization states of the photons.
\begin{center}
\begin{figure}[t]
\includegraphics[width=.45\textwidth]{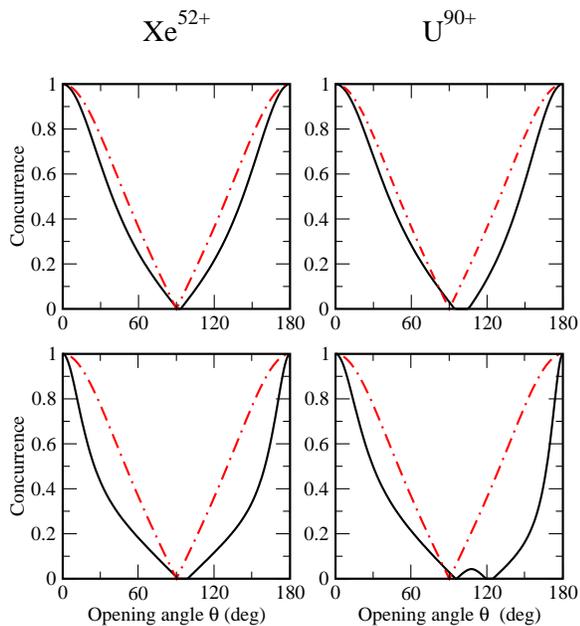}
\caption{(Color online) Concurrence of two photons emitted in the $1s_{1/2} \, 2s_{1/2} \; {}^3S_{1} \to 1s_{1/2}^2 \; {}^1S_0$ decay of helium--like xenon and uranium ions. Results of the $E1E1$ dipole approximation (dash-dotted line) and the rigorous relativistic theory (solid line) are shown, for two relative photon energies $\eta = 1/10$ (upper panel) and $\eta = 1/4$ (lower panel).
}
\label{Fig5}
\end{figure}
\end{center}
\subsection{Helium--like ions}
\label{sub_sec_helium}

\subsubsection{$1s_{1/2} \, 2s_{1/2} \; {}^1S_0 \rightarrow 1s_{1/2}^2 \; {}^1S_0$ transition}

In contrast to the $2s_{1/2} \to 1s_{1/2}$ decay of one--electron systems, the $1s_{1/2} \, 2s_{1/2} \; {}^1S_0 \to 1s_{1/2}^2 \; {}^1S_0$ transition in helium--like ions always proceeds between the \textit{pure}, $J = 0$, quantum--mechanical states (cf. Eq.~(\ref{initOP})). Therefore, the spin--state of the photon pair emitted in such a transition cannot be but pure at any energy sharing. The concurrence measure $C(\theta)$ of such a pure state calculated within the exact relativistic theory turns out to be almost identical to the dipole approximation (\ref{concket}). The deviation between both predictions does not exceed $1\%$, even for the heaviest helium--like ions and arises due to the higher, non--dipole terms in electron--photon interaction (\ref{R_interaction}). By comparing this prediction with the calculations performed in the previous section for hydrogen--like ions, we again argue that the reduction of the spin--entanglement between the photons emitted in the $2s_{1/2} \to 1s_{1/2}$ transition shall be mainly attributed to the loss of purity of ionic states.
\medskip

\subsubsection{$1s_{1/2} \, 2p_{1/2} \; {}^3P_0 \rightarrow 1s_{1/2}^2 {}^1S_0$ transition}

After the above, short discussion of the $1s_{1/2} \, 2s_{1/2} \; {}^1S_0 \to  1s_{1/2}^2 \; {}^1S_0$ transition, we turn now to explore quantum correlations in the $1s_{1/2} \, 2p_{1/2} \; {}^3P_0 \to 1s_{1/2}^2 \; {}^1S_0$ decay. As one can expect from the spin--state vector (\ref{ketP}), derived in leading order, ``electric and magnetic'' dipole approximation, these correlations should differ from those predicted for the $S \to S$ cases. Indeed, by inserting the vector (\ref{ketP}) into Eq.~(\ref{concket}), we obtain---within the dipole (E1M1) approximation---the concurrence measure as:
\begin{equation}
   \mathcal{C}(\theta, \eta)=\left| \frac{\Delta^2(\eta)\cos^4\frac{\theta}{2}-\Sigma^2(\eta)\sin^4\frac{\theta}{2}}
   {\Delta^2(\eta)\cos^4\frac{\theta}{2}+\Sigma^2(\eta)\sin^4\frac{\theta}{2}}
   \right| ~.
\label{concP}
\end{equation}
In contrast to the polarization entanglement (\ref{concket}) between the photons emitted in the $S \to S$ transitions, the concurrence turns out here to depend on the photons' energy sharing. To better understand such a dependence,  we display the entanglement of the photons emitted in the $1s_{1/2} \, 2p_{1/2} \; {}^3P_0 \to 1s_{1/2}^2 \; {}^1S_0$ decay of helium--like xenon and uranium ions in Fig.~\ref{Fig4}. 
Calculations are performed for two relative energies: $\eta$ = 1/10 and 1/4 both within the dipole approximation (\ref{concP}) and by using the exact theory which accounts for the higher multipole channels. As seen from the figure, both theoretical approximations predict maximal entanglement, $\mathcal{C} = 1$, for the parallel and back--to--back photon emission at \textit{any} energy sharing $\eta$; a feature that could be expected from the conservation laws. On the contrary, the ``critical'' opening angle $\theta_c$ at which the concurrence vanishes, $\mathcal{C}(\theta_c) = 0$, varies with the relative photon energy. By inspecting Eq.~(\ref{concP}) we find the following relation for this angle:
\begin{equation}
   \ds \tan^2\Big(\frac{\theta_c}{2}\Big) = \frac{\big|\Delta(\eta)\big|}{\big|\Sigma(\eta)\big|} \, .
   \label{minC}
\end{equation}
It follows from this expression that for any non--zero values of the functions $\Delta(\eta)$ and $\Sigma(\eta)$ there exists one single critical angle $\theta_c(\eta)$, as can also be seen, for example, from Fig.~\ref{Fig4}.

\medskip

If the photons, emitted in the $1s_{1/2} \, 2p_{1/2} \; {}^3P_0 \to 1s_{1/2}^2 \; {}^1S_0$ decay, carry away the same portion of energy, $\omega_1 = \omega_2$, 
the function $\Delta(\eta)$ turns out to be zero and Eq.~(\ref{minC}) cannot be applied for the 
determination of the critical angle $\theta_c$. 
%In this case, however, we can easily obtain the concurrence of the photons' state 
%directly from Eq.~(\ref{concP}): $\mathcal{C}(\theta, 0.5)$ = 1.
As can be seen from Eq.~(\ref{kketP}) and Eq.~(\ref{concP}), in this case the photons' state is maximally entangled (Bell state) for \textit{any} 
opening angle, i.e. $\mathcal{C}(\theta, 0.5)$ = 1. This behaviour differs from that of $1s_{1/2} \, 2s_{1/2} \; {}^1S_0 \to 1s_{1/2}^2 \; {}^1S_0$ as well as $2s_{1/2} \to 1s_{1/2}$ transitions for which no correlations appear at the opening angle $\theta = \pi/2$. In order to understand the reason for this difference, we shall return to Eqs.~(\ref{Mfi})--(\ref{A_decomposition}). By making use of these expressions and of the properties of the Wigner matrices we re--write the two--photon transition amplitude in terms of reduced matrix elements as:
\begin{equation}
   \ds\mathcal{M}_{fi}^{\vec k_1\vec k_2}(\lambda_1,\lambda_2) \propto
   (\lambda_1 + \lambda_2) \left( S_{E1 M1}(\omega) + S_{M1 E1}(\omega) \right) \, ,
   \label{Mfi_PtoS}
\end{equation}
where, for the case of equal energy sharing, $\omega_1 = \omega_2 = \omega$. It follows from Eq.~(\ref{Mfi_PtoS}) that apart from the conservation of the projection of the total angular momentum $J_{\rm tot}$, discussed in Section \ref{sub_sec_hydrogen}, an \textit{additional} selection rule arises for the $1s_{1/2} \, 2p_{1/2} \; {}^3P_0 \to 1s_{1/2}^2 \; {}^1S_0$ that forbids emission of the photons with opposite helicities. Together with the equal probabilities of the spin--states $\ket{+ +}$ and $\ket{- -}$, this selection rule implies the Bell state (\ref{kketP}) and, hence, maximal engagement of the photons' state.

\medskip
\subsubsection{Incoherent preparation of the ions: $1s_{1/2} \, 2s_{1/2} \; {}^3S_1 \rightarrow 1s_{1/2}^2 \; {}^1S_0$ transition}

Until now our discussion of the two--photon decay of helium--like ions was restricted to $J = 0 \to J = 0$ transitions. In this case, both initial and final ionic states are pure along the entire isoelectronic sequence, and, consequently, the two--photon states are pure as well.
In order to underline again the effect of the loss of purity on the quantum correlations, we study the two--photon decay of the unpolarized $1s_{1/2} \, 2s_{1/2} \; {}^3S_1$ state. In Fig.~\ref{Fig5}, we display the degree of spin--entanglement for the $1s_{1/2} \, 2s_{1/2} \; {}^3S_1 \to 1s_{1/2}^2 \; {}^1S_0$ transition in helium--like xenon and uranium ions. Again, the exact relativistic calculations are compared with the predictions of electric dipole ($E1E1$) approach for two relative energies $\eta$ = 1/10 and 1/4. As seen from the figure, the general behaviour of the measure $\mathcal{C}_{\, ^3S_0 \to ^1S_0}$ is very similar to that of the $1s_{1/2} \, 2s_{1/2} \; {}^1S_0 \to 1s_{1/2}^2 \; {}^1S_0$ and $2s_{1/2} \to 1s_{1/2}$ transitions. Namely, the concurrence changes from $\mathcal{C}_{\, ^3S_0 \to ^1S_0} = 1$ for the parallel photon emission down to zero at $\theta = \pi/2$ and back to a maximum entanglement for $\theta = \pi$. Similar to the discussion in Section \ref{sub_sec_hydrogen}, this can be easily understood if one applies again the momentum projection selection rules and Eq.~(\ref{A_Pi2}). In contrast to the $1s_{1/2} \, 2s_{1/2} \; {}^1S_0 \to 1s_{1/2}^2 \; {}^1S_0$ and $2s_{1/2} \to 1s_{1/2}$ transitions, however, the degree of entanglement $\mathcal{C}_{\, ^3S_0 \to ^1S_0}$ drops down much faster for the forward $0 < \theta < \pi/3$ and backward $2\pi/3 < \theta < \pi$ angles; an effect that can be understood if we remember that the initial ionic state is prepared in an unpolarized (mixed) state.

\medskip

As one can see from Fig.~\ref{Fig5},  spin entanglement for the $1s_{1/2} \, 2s_{1/2} \; {}^3S_1\to1s_{1/2}^2 \; {}^1S_0$ transition is very sensitive to higher multipoles in the electron--photon interaction. This is a direct consequence of a strong suppression of the $E1E1$ decay channel caused by the symmetry properties of the multi--photon systems as described by Bose statistics (see Refs.~\cite{PRA69,PRL83,SSR60,PR77} for further details). The non--dipole contributions become more significant for heavier ions and with increasing energy sharing $\eta$, $0 <\eta < 0.5$, and result in an asymmetric shift in the concurrence.

\section{Summary}
\label{sec_summary}

In summary, the two--photon decay of few--electron ions has been investigated within the framework of density matrix and second--order perturbation theory. Special attention in our study has been paid to the quantum correlations between the spin states of the emitted photons.
By making use of the non--relativistic dipole model, we derived a simple analytical expression for such spin--entanglement if observed in $2s_{1/2} \to 1s_{1/2}$ and $1s_{1/2} \, 2s_{1/2} \; {}^1S_0 \to 1s_{1/2}^2 \; {}^1S_0$ transitions in hydrogen and helium--like ions, respectively. By comparing predictions of the dipole approximation with the fully--relativistic calculations, we were able to explore the influence of the relativistic effects on the photon polarization properties. In particular, we observed a reduction of entanglement that becomes more sizable for high--$Z$ systems and that can be attributed to the loss of purity of the two--photon spin states induced by higher multipole contributions and the relativistic contraction of the wave--functions.

\medskip

Beside the well--established $2s_{1/2} \to 1s_{1/2}$ and $1s_{1/2} \, 2s_{1/2} \; {}^1S_0 \to 1s_{1/2}^2 \; {}^1S_0$ transitions, entanglement studies were also performed for the $1s_{1/2} \, 2s_{1/2} \; {}^3S_1 \to 1s_{1/2}^2 \; {}^1S_0$ and $1s_{1/2} \, 2p_{1/2} \; {}^3P_0 \to 1s_{1/2}^2 \; {}^1S_0$ decays of intermediate and high $Z$ helium like ions. Based on the independent particle model, which is a good approximation for the analysis of the bound--bound transitions in heavy atomic systems, we found that the concurrence is very sensitive to the relative energy of emitted photons, as well as to the higher multipole contributions to the electron--photon interaction. The strongest non--dipole effects have been identified for the $1s_{1/2} \, 2s_{1/2} \; {}^3S_1 \to 1s_{1/2}^2 \; {}^1S_0$ two--photon transition for which the $E1E1$ decay channel is forbidden due to symmetry properties of the system.

\medskip

Our theoretical analysis, performed for the intermediate and high $Z$ domain, underlines the importance of detailed knowledge on the electronic structure of ions (atoms) for a better understanding of two--photon entanglement. This is rather different from the earlier studies on the decay of light neutral atoms \cite{AsD82,PiK85,KlD97} where the quantum correlations between the photons could be predicted solely from angular--momentum conservation. In contrast, for heavy ions, entanglement properties of the photon pairs are governed by the complicated interplay between these conservation laws and relativistic as well as many--body effects.

\medskip

The spin quantum correlation studies reported in the present work will help to understand the outcome of the future $\gamma-\gamma$ coincidence measurements. Owing to the recent developments in x--ray polarization detectors, these measurements are likely to be performed at the GSI storage ring in Darmstadt. Apart from the analysis of  relativistic and quantum electrodynamics (QED) effects, parity--violation (PV) corrections to the photon spin--entanglement can be observed in such coincidence experiments. A theoretical analysis of these PV phenomena is currently under way. 

\section{Acknowledgements}
The authors would like to thank the organizers of the workshop EAS meeting 2009 in Riezlern where this work was initiated. A.S. and F.F. acknowledge the support from the Helmholtz Gemeinschaft under the project VH--NG--421. M.C.T. acknowledges financial support by Studienstiftung des deutschen Volkes and by DFG-Research Unit FG760.

%=========Bibliography========================================================

\newpage

\end{document}